\documentclass{sig-alternate-05-2015}

\usepackage{balance}

\usepackage{array}
\usepackage{booktabs}
\usepackage{pifont}
\usepackage{threeparttable}

\newcommand*\OK{\ding{51}}

\usepackage{blindtext}
\usepackage{graphicx}
\usepackage{amsmath}
\usepackage{amssymb}
\usepackage[mathlines,switch]{lineno}
\usepackage{breakurl}

\newenvironment{glist}[3]{
\begin{list}{#1}{\topsep 0.0mm
                        \partopsep 0.0mm
                        \setlength{\itemindent}{#3}
                        \parsep 0.0mm
                        \itemsep #2\baselineskip
                        \settowidth{\labelwidth}{#1}
                        \leftmargin 0.0mm
                        \addtolength{\leftmargin}{\itemindent}
                        \addtolength{\leftmargin}{\labelwidth}
                        \addtolength{\leftmargin}{\labelsep}}}{\end{list}}

\usepackage{blindtext}
\usepackage{graphicx}
\usepackage{amsmath}
\usepackage{amssymb}
\usepackage[mathlines,switch]{lineno}
\usepackage[numbers]{natbib}

\usepackage{amsfonts}

\usepackage{multirow}

\usepackage{slashbox}

\usepackage{units}

\usepackage{algorithm2e}

\usepackage[tight,footnotesize]{subfigure}

\usepackage{fixltx2e}
\usepackage{resizegather}

\begin{document}


\title{GRM: Group Regularity Mobility Model}

\author{Ivan O. Nunes, Clayson Celes, Michael D. Silva, Pedro O.S. Vaz de Melo, Antonio A.F. Loureiro\\
Department of Computer Science\\
Federal University of Minas Gerais\\
Brazil\\
\{ivanolive,claysonceles,micdoug,olmo,loureiro\}@dcc.ufmg.br
}

\date{5-8 July 2016}




\maketitle

\begin{abstract}
In this work we propose, implement, and evaluate GRM, a novel mobility model that accounts for the role of group meeting dynamics and regularity in human mobility. Specifically, we show that existing mobility models for humans do not capture the regularity of human group meetings which is present in real mobility traces. Next, we characterize the statistical properties of such group meetings in real mobility traces and design GRM accordingly. We show that GRM maintains the typical pairwise contact properties of real traces, such as contact duration and inter-contact time distributions. In addition, GRM accounts for the role of group mobility, presenting group meetings regularity and social communities' structure. Finally, we evaluate state-of-art social-aware protocols for opportunistic routing using a synthetic contact trace generated by our model. The results show that the behavior of such protocols in our model is similar to their behavior in real mobility traces.
\end{abstract}


\section{Introduction}\label{intro}

Mobility models have fundamental importance for mobile networking prototyping. They enable the generation of synthetic trajectories for mobile nodes in simulated environments, which can then be used to evaluate the performance of newly designed networking protocols. The validation of such protocols in real-world large scale experiments is often unfeasible due to the financial and operational limitations. On the other hand, synthetic models enable the rapid performance evaluation of networking protocols considering long-term evaluation periods, and larger number of network nodes.

In the past years, several mobility models were proposed with the goal of reproducing one or more statistical properties of the human mobility. Examples of such properties include human walks and displacements \cite{individual_mob,rhee2011levy}, the spatial regularity of human mobility \cite{PappalardoNature2015,wdm}, human trajectories and transportation \cite{mobhet,silva2015filling}, pairwise encounter patterns \cite{swim,slaw}, and also group mobility \cite{HongMSWIM1999,blakely2004structured,musolesi2006community}. 

The group mobility property is already considered a fundamental building block for mobility modeling \cite{survey1}. However, the existing group mobility models for humans focus on modeling groups that remain together throughout the whole simulation time. Therefore, such models are not representative of the statistical regularity of human interactions, i.e., groups of people that meet regularly. Recent studies \cite{icc,cruz2015recurring,starnini2014model,wcm} have shown that the regularity of group meetings, present in real-world traces, play an important role for content forwarding in mobile opportunistic networks.

Although earlier studies~\cite{wdm,swim,slaw} on mobility modeling have focused on reproducing the regularity of human contacts, those models only focus on reproducing the regularity of pairwise interactions, i.e., they only model contacts between two people, disregarding collective social interaction. In other words, such models do not account for the role of group meetings. This limitation is specially harmful to the validation of opportunistic forwarding protocols (e.g., DTN~\cite{zhu2013survey} and D2D protocols~\cite{asadi2014survey}), because the social-aware strategies \cite{bubble,D2Dbubble} have remarked themselves as the most effective for this types of protocols. As a direct consequence, none of the existing mobility models is qualified for evaluating the socially-aware approaches for opportunistic routing, since they do not completely capture the social regularity presented by human mobility.

Aiming at addressing the aforementioned issues, in this work we propose the Group Regularity Mobility (GRM) Model\footnote{GRM synthetic mobility traces containing 100, 1000, and 2000 mobile nodes are available together with a demo video of GRM working on top of The ONE Simulator~\cite{keranen2009one} at:\\ \textbf{\burl{https://www.dropbox.com/sh/792mi849nf3dvam/AAAR4RofaLBfoFaxmeONe-H4a?dl=0}}\\\\ GRM source code is available at:\\ \textbf{\burl{https://github.com/ivanolive/GRM}}}. GRM is the first mobility model to consider the role of group meetings and their regularity to simulate human mobility. We show that important real-world mobility properties, such as social community structure in the mobile network, group meetings regularity, and statistical patterns of inter-contact time and contact duration, are well retained by GRM. Moreover, we evaluate state-of-art socially-aware opportunistic forwarding protocols and show that their performances in synthetic traces generated by GRM are very similar to their performance in real-world traces. In summary, the contributions of the present work are threefold:
\begin{glist}{$\bullet$}{0}{0mm}
\topsep 0mm
\parskip 0mm
\partopsep 0mm
\parindent 0mm
\itemsep 0mm
\parsep 0mm
     \item We empirically show that the existing mobility models for opportunistic networking do not present group meetings regularity.
     \item We use the two publicly available real mobility data-sources with the largest scale, in time duration and number of nodes, to leverage and characterize the group meetings properties and design GRM: the first mobility model to capture the role of group mobility and its regularity.
     \item We evaluate GRM and show that it has the same statistical properties of real mobility traces, considering inter-contact time, contact duration, social community structure, and group mobility regularity. We also show that the state-of-art socially aware opportunistic forwarding protocols present the same performance in GRM as they do in real mobility traces.
 \end{glist}

This paper is organized as follows. In Sec.~\ref{related}, we review related research efforts highlighting their contributions to mobility modeling. In Sec.~\ref{motivation}, we evaluate the state-of-art mobility models and show that they are not representative of the group mobility regularity which is present in real-world mobility traces. In Sec.~\ref{grm}, we describe the GRM model, providing an in-depth discussion of each one of the building blocks used in the model design. In Sec.~\ref{eval}, we statistically characterize and compare GRM to real-world traces. The results confirm that GRM performs accordingly. In Sec.~\ref{forwarding},  we emulate the state-of-art socially aware protocols for opportunistic forwarding in synthetic traces generated by GRM. Finally, in Sec.~\ref{conclusion}, we present the final remarks and future work.

\section{Related Work}\label{related}


Group mobility is already considered a fundamental building block for mobility modeling \cite{survey1}. However, the existing group mobility models \cite{survey5} focus on modeling groups, which remain together throughout the whole simulation time. On the other hand, mobility models that model the regularity of human contact patterns \cite{wdm,slaw,kosta2014large} only consider pairwise contacts, ignoring the fact that human social contacts often happen in groups, involving more than two entities. 

Existing group mobility models are restricted to represent nodes that move together as clusters. For example, Reference Point Group Mobility (RPGM) \cite{HongMSWIM1999} and Reference Velocity Group Mobility (RVGM) \cite{WangICC2002} are variants of random models for group mobility. In both, people are organized in groups based on their logical relationships. Each group contains one leader, and the members of a group follow their leader. These mobility models are based on certain properties of movement, such as speed, direction, and acceleration, and do not exhibit the typical contact properties of real-world human mobility. Therefore, such models are not able to reproduce social structures and statistical properties of real-world mobility~\cite{survey2}.

In recent years, some studies have focused on modeling human mobility using spatial and temporal statistical patterns that were observed in real-world mobility traces. Lee et {al.}\ \cite{slaw} presented the Self-similar Least Action Walk (SLAW) mobility model that captures the following properties: truncated power-law distributions of flights, pause-times and inter-contact times, attractive force to more popular places, and heterogeneously defined areas of individual mobility. The model uses these features to represent the mobility of people who share ``common gathering places'', i.e., places that most people visit during their daily lives.

\begin{table}[]
    \caption{Opportunistic networking properties in each mobility model: Groups of Nodes (GN), Inter-Contact Time (ICT), Contact Duration (CD), Displacements (D), Social Context (SC), and Group Meetings Regularity (GMR).}
    \label{tab:properties}
 \begin{center}
   \begin{threeparttable}
    \begin{tabular}{|l|c|c|c|c|c|c|}
    \hline 
    \backslashbox{Model}{Property} &
    GN & ICT & CD & D & SC & GMR
        \\ \hline \hline
    RPGM~\cite{HongMSWIM1999}  & \OK & --- & --- & --- & --- & --- \\ \hline
    RVGM~\cite{WangICC2002}    & \OK & --- & --- & --- & --- & --- \\ \hline
    CMM~\cite{musolesi2007}    & --- & \OK & \OK & --- & \OK & --- \\ \hline
    WDM~\cite{wdm}             & --- & \OK & \OK & \OK & --- & --- \\ \hline
    SLAW~\cite{slaw}           & --- & \OK & \OK & \OK & --- & --- \\ \hline
    HCMM~\cite{boldriniComCom2010}   & --- & \OK & \OK & \OK & \OK & ---\\ \hline
    SWIM~\cite{swim}           & --- & \OK & \OK & \OK & \OK & --- \\ \hline
    GRM (our model)           & \OK & \OK & \OK & \OK & \OK & \OK \\ \hline
    \end{tabular}
    \begin{tablenotes}
        \item \small Note: ''\OK'' if the model satisfies the property, "---" otherwise.
    \end{tablenotes}
    \end{threeparttable}
  \end{center}

\end{table}

In Small World in Motion (SWIM) \cite{swim}, Kosta el {al.}\ presented a mobility model based on the intuition that people go more often to nearby or popular places. This intuition is supported by Gonzalez et {al.}\ \cite{nature} observation of spatial and temporal regularity in human movement. SWIM assigns each node with a home location and computes a visitation probability to each possible destination in the simulation space. The visitation probability in computed according to (i) the popularity of each possible destination, and (ii) the distance of each location to the node's home location (closest destinations have higher probabilities).

SLAW and SWIM produce inter-contact time and contact duration distributions that follow the ones found in real mobility traces. However, both models consider only pairwise contacts, ignoring group mobility or any relationships between more than two nodes.

\begin{figure*}[!t]
\centering
  \subfigure[MIT (Real)]{\includegraphics[width=3in]{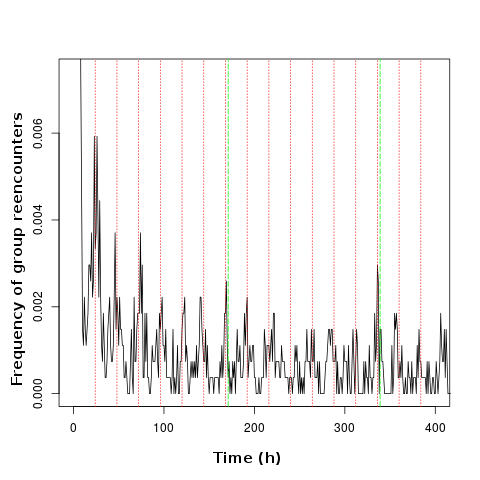}\label{peri_mit}}
  \subfigure[Dartmouth (Real)]{\includegraphics[width=3in]{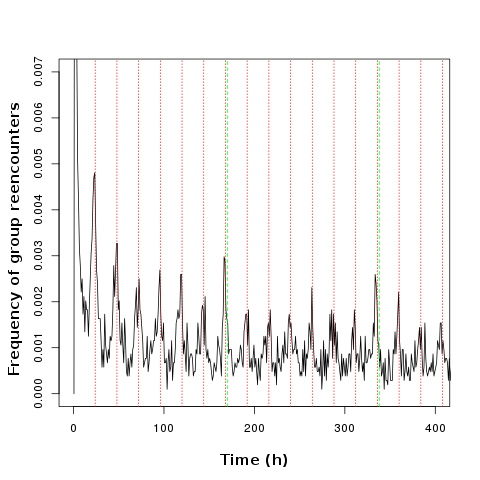}\label{peri_dartmouth}}
  \subfigure[SWIM (Synthetic)]{\includegraphics[width=3in]{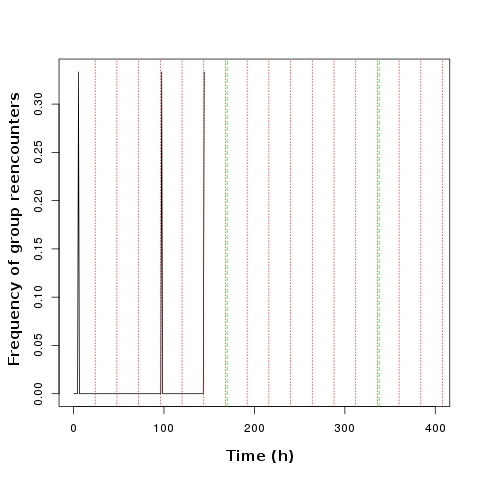}\label{peri_swim}}
  \subfigure[WDM (Synthetic)]{\includegraphics[width=3in]{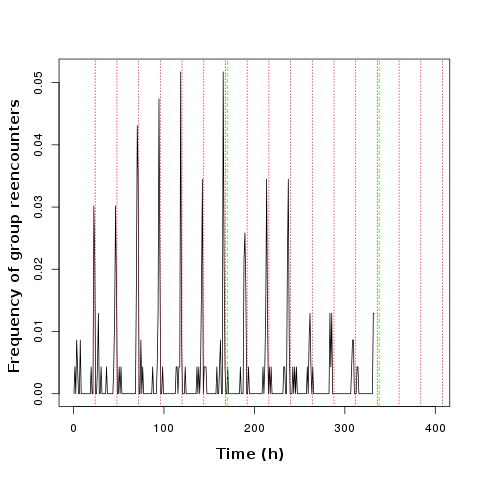}\label{peri_wdm}}
\caption{Comparison of group meetings regularity in real and synthetic mobility traces}\label{periodicity}
\end{figure*}

Musolesi and Mascolo \cite{musolesi2007} proposed the community based mobility model (CMM) founded on social network theory. The model receives a social network as input and applies a community detection algorithm to define the nodes' movements according to the social ties between them. The intuition is that nodes go to places with higher social attraction. Boldrini and Passarella \cite{boldriniComCom2010} presented the Home Cell Mobility Model (HCMM), which is an evolution of the CMM. HCMM focus on ideas of social and location attractions. Similarly to SWIM, the HCMM adopted the concept of home location and the nodes movements are conditioned by their social relationships. Moreover, nodes go to few places more often and these places are not far from their homes. In these models, the community structure is forced into the nodes' mobility to generate social context. In our model, on the other hand, the community structure emerges naturally from the regularity of group meetings, and from the dynamically defined group composition, as it happens in real world.

With the goal of modeling daily behavior, Ekman et {al.}\ \cite{wdm} have introduced Working Day Movement Model (WDM). WDM simulates daily routines of people considering their commutes between home and workplace. WDM presents human mobility regularity, but, as we show in Sec.~\ref{motivation}, it is not representative of real-world group mobility.

In summary, Table \ref{tab:properties} presents the fundamental properties for opportunistic networking and their presence/absence at each mobility model. GRM is an evolution of the aforementioned models, including all of their properties and also group meetings regularity. In Sec. \ref{motivation}, we extend this discussion by providing empirical evidence that current state-of-art mobility models are not representative of the statistical regularity of human interactions when such interactions involve groups.

\section{Group Mobility: Real-world vs. Synthetic Models}\label{motivation}

\begin{figure}[!t]
\centering
\includegraphics[width=3in]{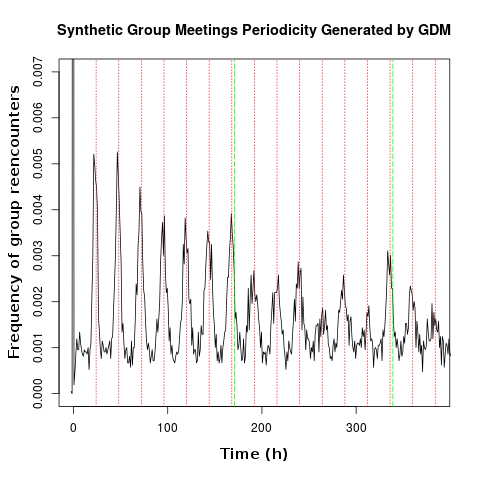}
\caption{Group meetings regularity in our model}\label{peri_grm}
\end{figure}

In this section, we compare some of the state-of-art and widely used synthetic mobility models with real mobility traces, with the goal of verifying if group meetings' regularity properties are captured by such synthetic models. Specifically, we want to verify if such models capture group re-encounters and their evolution over time, i.e., groups' dynamics, to be able to decide if they are representative of real-world group mobility.

To that purpose, we apply the methodology for detecting and tracking mobile groups, proposed in \cite{icc}, to the real mobility traces, MIT and Dartmouth. MIT \cite{mit} and Dartmouth \cite{dartmouth} traces are proximity contact registers containing 80 and 1200 users respectively. In the MIT trace, monitored users reside in two university buildings and were monitored for almost one year. Contacts were registered when two users were less than 10 meters apart. The Dartmouth trace registered contacts of the students in a university campus for two months using Wi-Fi connectivity logs. To the best of our knowledge, Dartmouth is the largest scale and publicly available contact dataset in terms of number of monitored nodes. On the other hand, the MIT trace is the publicly available dataset which monitored its users for the largest period of time.

Figs.~\ref{peri_mit} and \ref{peri_dartmouth} show the P.D.F. of group re-meetings along the time for the real-world traces. In both real mobility traces, we can verify the presence of periodicity in groups' re-encounters. In both of them, the mass of probability is concentrated in peaks around the red dotted lines, which represent periods of 24 hours. We also observe in both Figs.~\ref{peri_mit} and \ref{peri_dartmouth} that higher peaks are presented around the green dashed lines, which represent periods of seven days. This pattern in the group re-meetings' P.D.F. shows that group meetings present daily and weekly periodicity. It is noteworthy that such pattern happens in both real traces, even though they are from different places, have different number of nodes, and used different data collection methods. Next, we leverage three widely used state-of-art synthetic mobility models to verify if they represent well the role of social groups to mobility.

The SWIM mobility model \cite{swim} generates synthetic small worlds, which preserve the pairwise contact duration and inter-contact times statistical distributions as they are observed in real mobility traces. The SLAW mobility model \cite{slaw} captures several significant statistical patterns of human mobility, including truncated power-law distributions of human displacements, pause-times and pairwise inter-contact times, fractal way-points, and heterogeneously defined areas of individual mobility. The Working Day Movement (WDM) synthetic model \cite{wdm} captures these same statistical properties of contact durations and inter-contact times as SWIM and SLAW. In addition to those properties, WDM aims at capturing the daily regularity of human movements, i.e., how human routines affect their mobility.

As we did for the real traces, MIT and Dartmouth, we have applied the same group detection and tracking methodology to the contact traces generated by these three synthetic models. Figs.~\ref{peri_swim} and~\ref{peri_wdm} present the results for the SWIM and WDM models, respectively.

The contact trace generated by the SWIM model (Fig. \ref{peri_swim}) do not present any group meeting regularity. Out of the detected groups, only three group re-meetings were registered in a period of 15 days. The result for the contact trace generated by the SLAW model presented an analogous behavior, i.e., no regularity in group meetings. This behavior is explained by the fact that such models were designed to be representative of the statistical properties of pairwise contacts only, without considering that human contacts often involve more than two peers. These models look only at pairwise contacts, disregarding group meetings.

In the WDM trace (Fig. \ref{peri_wdm}) we can observe that group re-meetings happen precisely in periods of 24 hours and with much higher frequencies than in real mobility traces. This behavior is observed because WDM firstly defines a set of places, called offices, and then distributes nodes to transition between pre-defined subsets of offices with daily periodicity. Therefore, nodes with intersections in their lists of offices will always form groups with exaggerated meeting regularity.


By analyzing the group meetings regularity of the synthetic models, we conclude that none of them represents well the group mobility patterns. Therefore, in Sec.~\ref{grm}, we propose GRM, a group dynamics aware mobility model, which incorporates the statistical properties of group meetings regularity.
In contrast with the existing mobility models, GRM is able to produce mobility traces that present group meetings regularity, as shown in Fig.~\ref{peri_grm}.

\section{The GRM Model}\label{grm}

\begin{figure}[!t]
\centering
\includegraphics[width=3in]{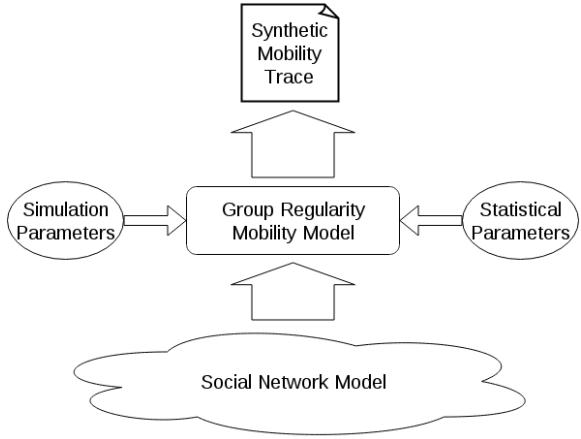}
\caption{GRM Model Framework}\label{GDM}
\end{figure}

\begin{figure}[!t]
\centering
  \subfigure[Dartmouth (Real Trace)]{\includegraphics[width=0.49\columnwidth]{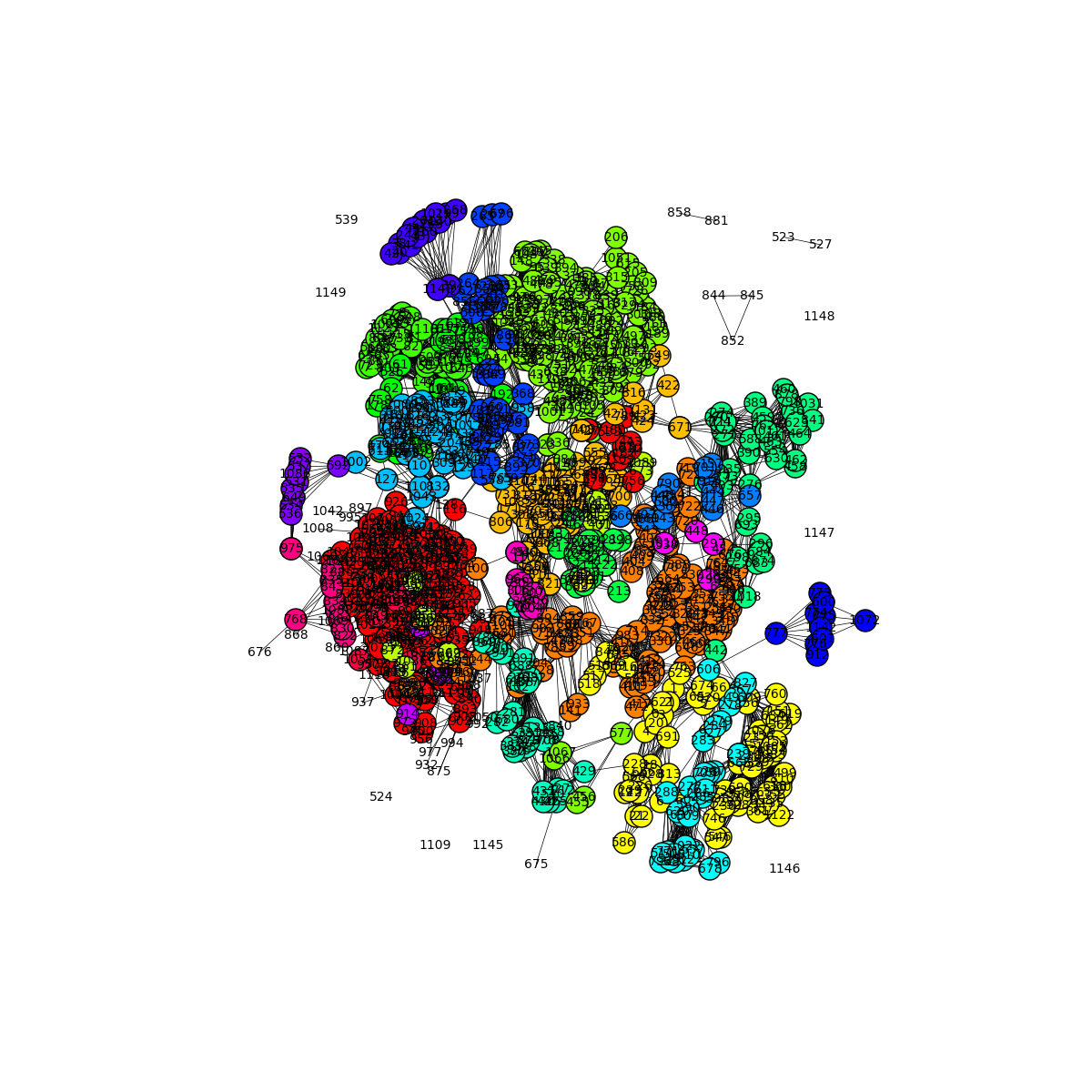}\label{dartmouth_net}}
  \subfigure[Barabasi-Albert (Synthetic Network Model)]{\includegraphics[width=0.49\columnwidth]{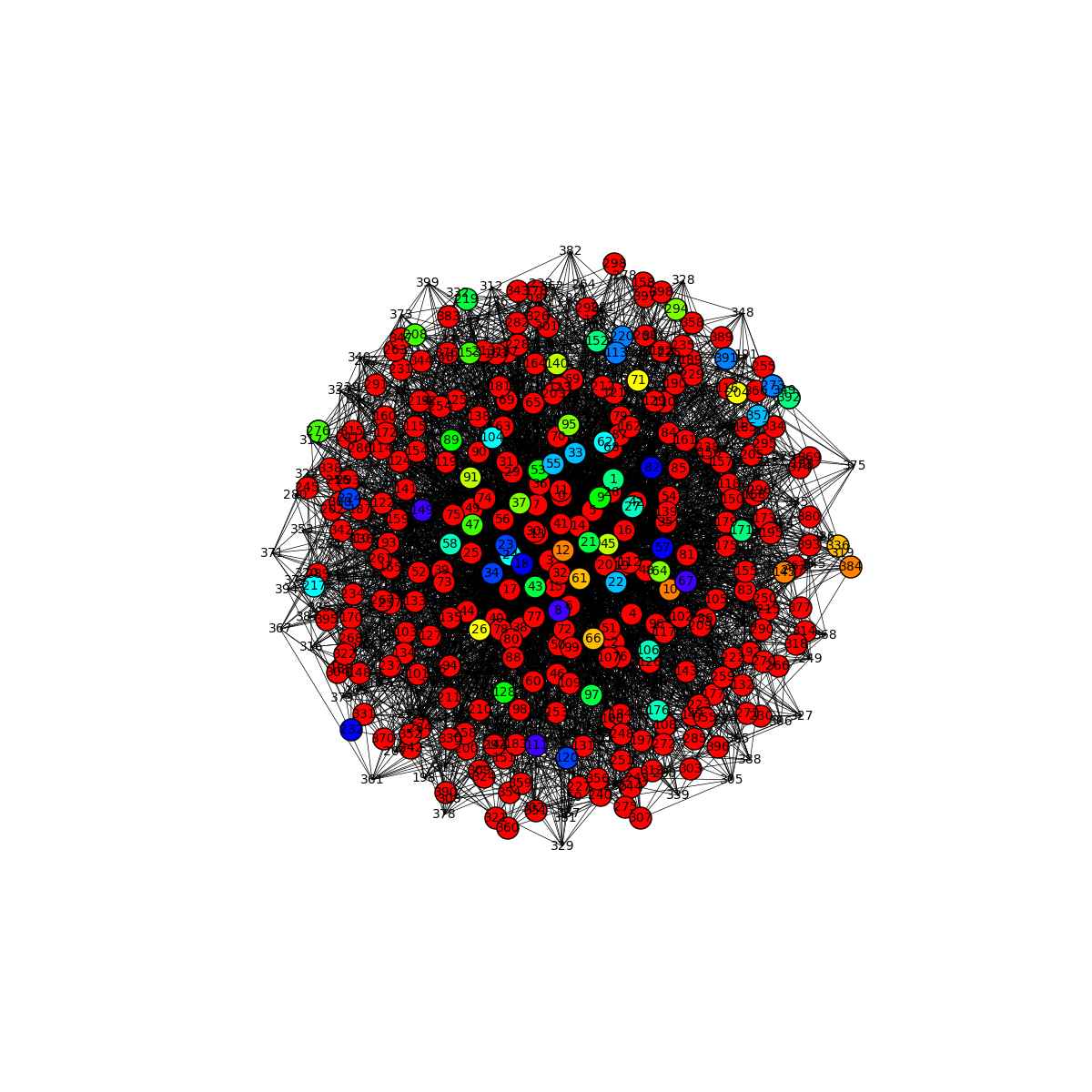}\label{barabasi_net}}
  \subfigure[Ex1 of Gaussian Clustering Model (Synthetic Network Model)]{\includegraphics[width=0.49\columnwidth]{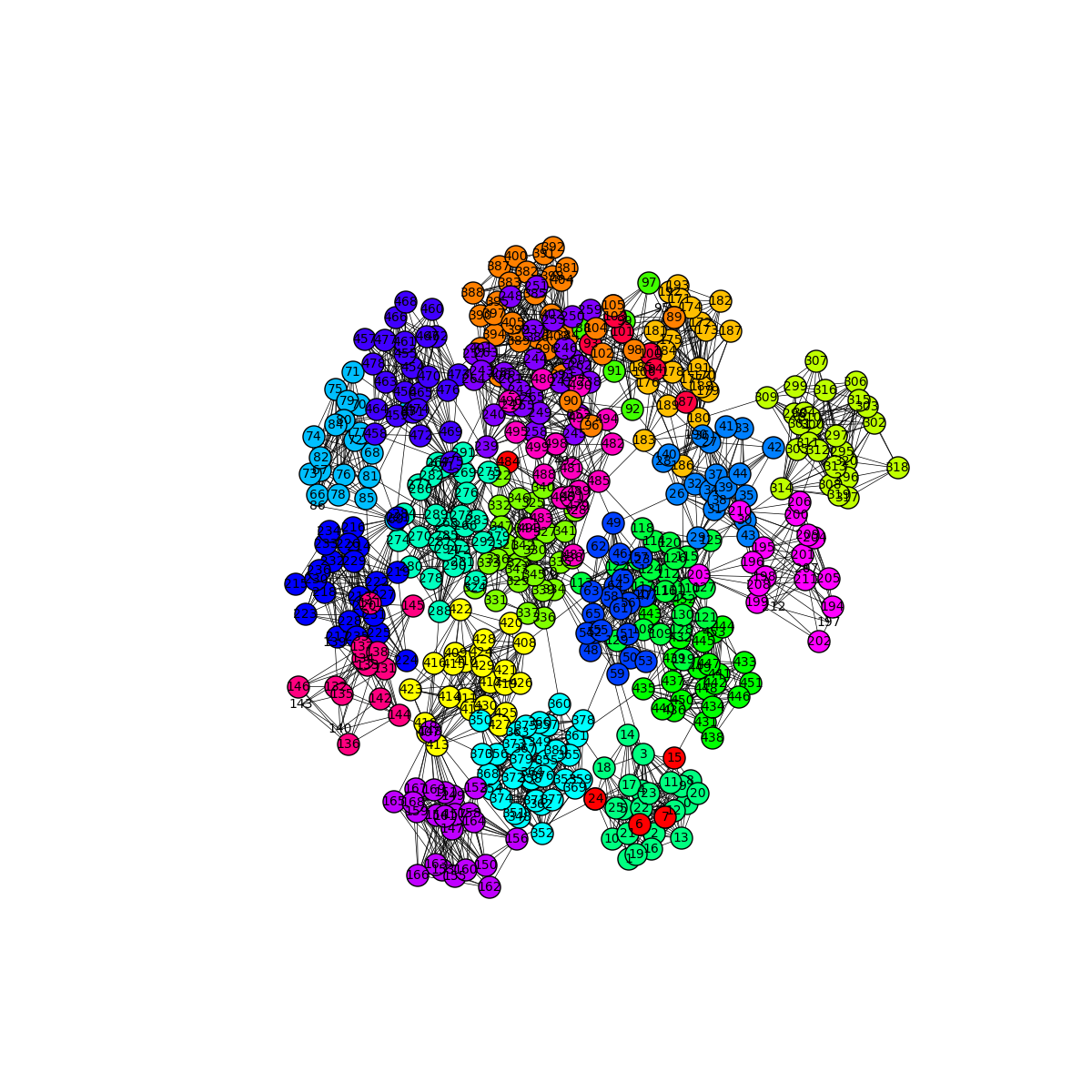}\label{gaussian_net1}}
  \subfigure[Ex2 of Gaussian Clustering Model (Synthetic Network Model)]{\includegraphics[width=0.49\columnwidth]{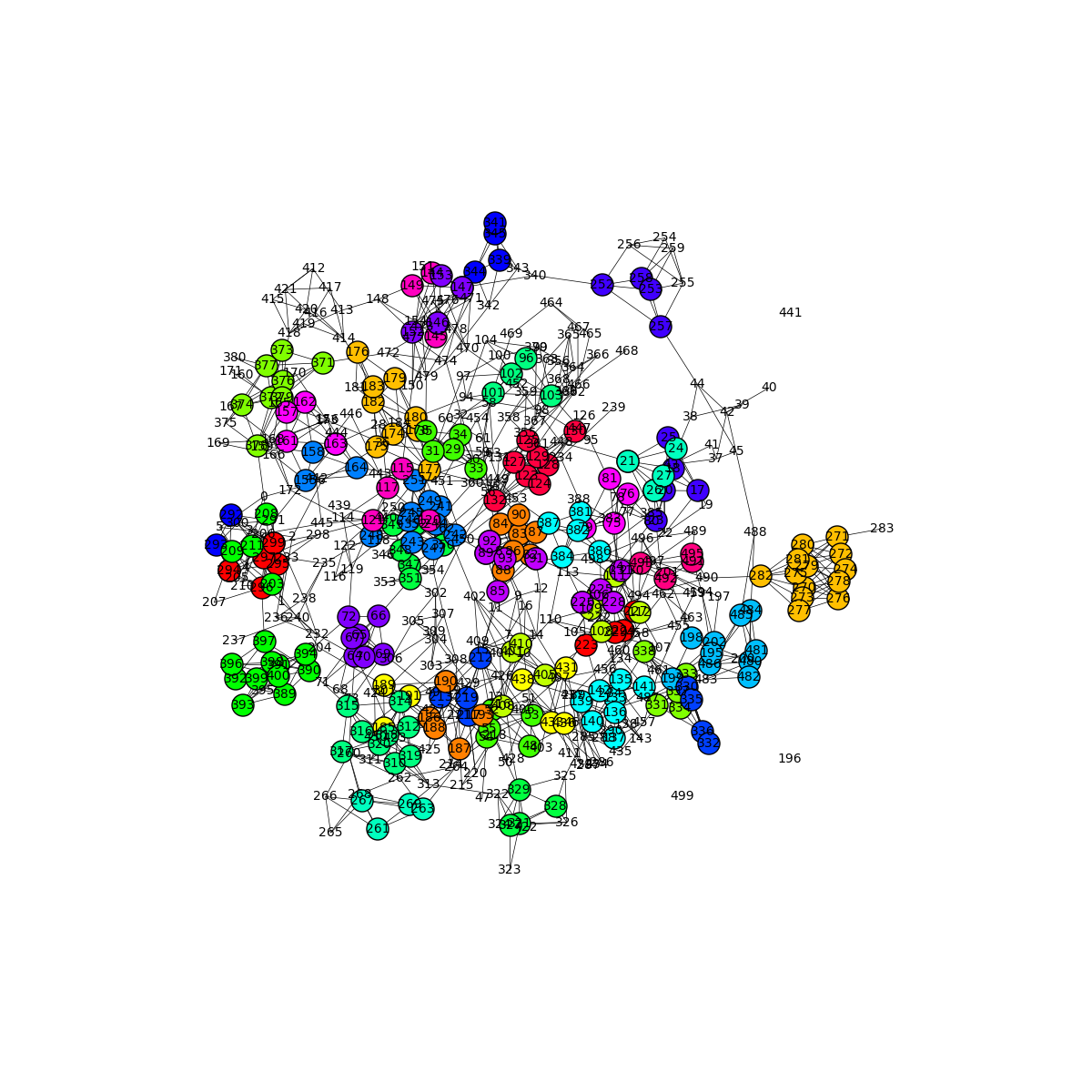}\label{gaussian_net2}}
\caption{Examples of social network models which can be given as input to GRM. Different colors denote communities detected in such networks}\label{graphs}
\end{figure}

In this section, we describe GRM in details. We go over each of the building blocks that are contained within the model. Fig.~\ref{GDM} presents the GRM functioning framework. GRM receives as input a social network, which can be a real social network, given as input by the user, or generated by a synthetic social network model. GRM implementation has native support for several social network models including Barabasi-Albert~\cite{barabasi1999emergence}, Gaussian Clustering~\cite{brandes2003experiments},  Caveman~\cite{watts1999networks}, and Random Partition Graph~\cite{fortunato2010community} models. The social network is used to define which nodes will be present at each group meeting event, i.e., the groups' structures, as discussed later on. The idea of providing the social network as an input for the model is to give flexibility for the mobility modeling and the social network modeling to evolve separately. GRM will adapt to any social network given as input and produce a mobility trace as output. Fig.~\ref{graphs} exemplifies social networks that are supported by GRM.

In addition to the social network, GRM receives a set of simulation configurations, which comprise, for instance, the size of the simulated area, the simulation duration, the number of nodes, and the number of groups. Finally, it also receives a set of statistical parameters, which are the parameters for the statistical distributions contained in the model. Such statistical parameters vary in different real mobility traces, depending on the scenario. Therefore, the values of these parameters can be given as input to the model directly, or via automated extraction from existing real-world mobility traces (see Sec.~\ref{auto} for details on the latter), allowing GRM to mimic the scenario and mobility behavior of a given real-world trace. The synthetic traces generated by GRM are fully compatible and are ready to run on top of the ONE simulator~\cite{keranen2009one}.

The summary for the notation we will use to describe GRM is provided in Table~\ref{notation}.

\renewcommand{\arraystretch}{1.5}

\begin{table}[]
\centering
\caption{Notation summary}
\label{notation}
\begin{tabular}{|l|p{6cm}|}
\hline
Notation                                 & Description                                                                                                                     \\ \hline \hline
$T$                                      & The trace duration        \\
\emph{NodesSet}               &  The set of all network nodes\\
$G_i$                                    & The $ith$ group of nodes in the trace      \\
$||G_i||$               &  The number of group members of $G_i$ \\
$T_{G_i}$                                & The existence period of group $G_i$ \\
$\mu_{G_i}$   & The average inter-meeting time for $G_i$   \\
$\emph{Meeting}_{G_i}(t)$                                & The time for the $t^{th}$ meeting of $G_i$  \\ 
$\emph{Dur}_{G_i}$               &  The duration of $G_i$ group meetings\\

$u \sim U(a,b)$         &  $u\in\mathbb{R}$ is a value randomly selected with uniform probability in the interval $[a,b]$                             \\
$\eta \sim N(\mu,\sigma^2)$         &  $\eta\in\mathbb{R}$ is a value randomly selected with a Gaussian distribution of mean $\mu$ and variance $\sigma^2$                             \\
$\rho \sim PL(\alpha,\beta)$        & $\rho\in\mathbb{R}$ is a value randomly selected with a truncated power law distribution with exponent $\alpha$ and the exponential cut-off value $\beta$ \\
$P_{att}[U_j,G_i]$        & The probability of user $U_j$ attending to a meeting of group $G_i$  \\
$P_{\rm{place}}(C_j,G_i)$        & The probability of a meeting of group $G_i$ to happen at the $C_j$ cell  \\

\hline

\end{tabular}

\end{table}

\subsection{Group Meeting Times}

To properly design a group regularity mobility model, there must be a representative statistical model for group meeting times. Due to group meetings' periodicity, presented in Fig. \ref{periodicity}, we model group meeting times as follows.

Each group $G_i$ in the model receives an average inter-meeting time, $\mu_{G_i}$. The value of $\mu_{G_i}$ is randomly generated according to a power-law distribution with exponential cut-off. This way of generating $\mu_{G_i}$ is based on the fact that inter-contact times of real mobility traces follow this distribution (as discussed in Secs.~\ref{intro} and \ref{related}). The power-law exponent ($\alpha_{\rm{gmt}}$) and the exponential cut-off value ($\beta_{\rm{gmt}}$) are statistical parameters given as input to the model. Then, a series of meeting times for group $G_i$ is recursively generated with Gaussian inter-meeting times, as in Eq.~\ref{rec_meetings}:

\begin{equation}\label{rec_meetings}
\begin{gathered}
 \mu_{G_i} \sim PL(\alpha_{\rm{gmt}}, \beta_{\rm{gmt}}) \\
 \emph{Meeting}_{G_i}(t) =
  \begin{cases}
    u \sim U(0,T)       & \text{if } t = 0\\
    \emph{Meeting}_{G_i}(t-1) + \eta \sim N(K\times\mu_{G_i},\sigma^2) & \text{if } t > 0\\
  \end{cases}
\end{gathered}
\end{equation}

In the simulation, each group $G_i$ has its own $\mu_{G_i}$. The variance $\sigma^2$ is a simulation parameter for all groups which allows higher or lower variation on the group meetings punctuality, according to the Gaussian distribution variance properties.

Following the recursive equation for group meetings generation, each group will then have its set of meetings determined as:
\begin{equation}\label{bla}
\begin{split}
\bigcup_{j=0}^{\lceil\frac{T_{G_i}}{K\times\mu_{G_i}}\rceil}{\emph{Meeting}_{G_i}(j)}, 
\end{split}
\end{equation}
where $T_{G_i}$ denotes the period of time throughout which the group $G_i$ will exist. GRM considers that each group $G_i$ has its own regularity factor, which is represented by the scale factor $K$ in Eq.~\ref{rec_meetings}. For instance, most of the groups with $K = 24h$ will usually meet every 24, 48, or 72 hours, following the power-law probability function of $\mu_{G_i}$. $K$ is a multiplier that will generate the periodical behavior of real traces, depicted in Fig.~\ref{periodicity} of Sec.~\ref{motivation}, while the value of $\mu_{G_i}$, generated by a truncated power law, will generate statistically representative inter-contact times.

Since each group has its own $K$ value, the distribution for the values of $K$ is given to the model as a simulation parameter. An example would be: ``The simulation will have 500 groups. Seventy percent of these groups will have $K = $24h, 15\% will have $K = $7 days, and 15\% $K = 6$h''. In Sec.~\ref{eval} we show that this example of configuration for the $K$ distribution generates group re-meetings that are very similar to the ones observed in the MIT and Dartmouth traces.

\subsection{Group Meetings Durations}

Since we have generated the group meeting times, we now must define the duration of a group meeting, i.e., the time that the involved nodes will spend together. To do so, we inherit the findings of previous studies (as discussed in Secs.~\ref{intro} and \ref{related}), which show that contact durations follow truncated power laws. Therefore, as we did for $\mu_{G_i}$ in Eq.~\ref{rec_meetings}, we define the meeting durations as:
\begin{equation}\label{dur_meetings}
\begin{split}
\emph{Dur}_{G_i} \sim PL(\alpha_{\rm{dur}},\beta_{\rm{dur}}),\\
\end{split}
\end{equation}
where $\alpha_{\rm{dur}}$ and $\beta_{\rm{dur}}$ are statistical parameters of GRM.

\subsection{Groups' Structure and Social Context}\label{structure}

\begin{figure}[!t]
\centering
  \subfigure[$\alpha= 2.24;\beta = 30.4$. Average group size of 6.06]{\includegraphics[width=0.49\columnwidth]{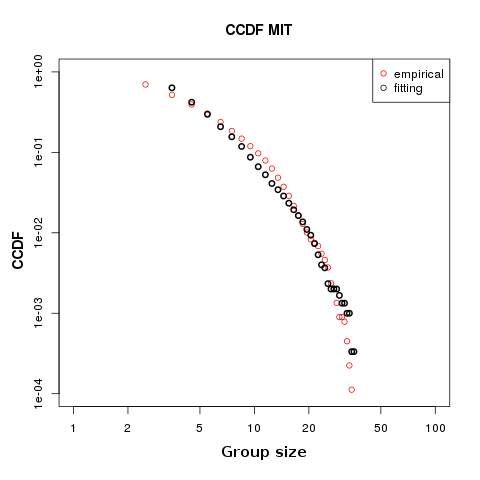}\label{mit_g_size}}
  \subfigure[$\alpha= 2.42; \beta= 54.6$. Average group size of 6.96]{\includegraphics[width=0.49\columnwidth]{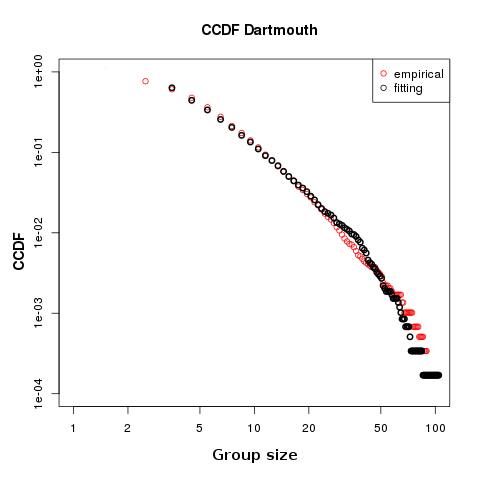}\label{dt_g_size}}
\caption{Group sizes: empirical data from MIT and Dartmouth traces and their fitting to data generated by power laws of exponents $\alpha$ and exponential cuts $\beta$.}\label{g_size}
\end{figure}

So far, we have defined how to generate group meeting times and their durations. Now we discuss how we define which nodes will be at each meeting, i.e., the groups' compositions. The first step to define group structures is to verify the group sizes in real mobility traces. In Fig.~\ref{g_size}, we show that group sizes in the MIT and Dartmouth traces follow power laws with exponential cuts with different parameters. Therefore, the number of group members in $G_i$ is defined as:
\begin{equation}\label{sizes}
\begin{split}
||G_i|| \sim PL(\alpha_{\rm{size}},\beta_{\rm{size}}),\\
\end{split}
\end{equation}
where $\alpha_{\rm{size}}$ and $\beta_{\rm{size}}$ are the last couple of statistical parameters of GRM.

GRM defines the network nodes that will compose a given group $G_i$ using the size $||G_i||$, defined by Eq.~\ref{sizes}, and a probabilistic snowball sampling algorithm \citep{snowball}. To do so, a node $n$ is randomly selected, with uniform probability, from the set of network nodes. The snowball algorithm randomly selects a set of neighbors of $n$. Next, it select a random set of the neighbors of the neighbors of $n$, and so on, until the set of selected nodes reach the predetermined size $||G_i||$. The selected set of nodes will compose the group $G_i$. The snowball sampling is performed in the inputted social network (see Fig.~\ref{graphs} for examples), thus the snowball sampling preserve the social context of such network. In summary, the structural composition of a group is defined as:

\begin{equation}\label{eq7}
\begin{gathered}
\emph{Node}_n = U(\emph{NodesSet\/})\\
\emph{Members}_{G_i} = \textrm{Snowball}(\emph{Node}_n,||G_i||,\emph{SocialGraph\/})
\end{gathered}
\end{equation}

At this point, it is worth to emphasize that, as it happens in reality, one node may participate of several social groups. In addition, the number of possible group structures is combinatorial in relation to the number of nodes. In practice, the number of groups detected in a real mobility trace is bigger than the number of nodes. For instance, around 5000 different groups were detected in the Dartmouth trace that monitors only 1200 nodes. 

Also, in reality, it is not reasonable to expect every node always attend to every meeting of a given group. In GRM, each user $U_j$, that is a member of the group $G_i$, receives a probability $P_{att}[U_j,G_i]$ of attending to a $G_i$ meeting as:

\begin{equation}\label{eq8}
\begin{split}
P_{\rm{att}}[U_j,G_i]= \frac{\textrm{Known}(\emph{User}_j,G_i,\emph{SocialGraph})}{||G_i||}
\end{split}
\end{equation}

The intuition behind the $P_{\rm{att}}$ probability is that people have higher probability to attend to meetings of social groups in which they know more nodes. The \emph{Known} function returns the number of nodes in $G_i$ that have social edges with $U_j$ in the inputted social network \emph{SocialGraph}.  

Using such modeling, each social group in the trace will have a different composition at each meeting, but, at the same time, maintaining most of its structure throughout all of its meetings. Such behavior is also presented in social relationships of real life~\cite{icc}.

\subsection{Mobility and Meeting Places}

The final step of GRM is to generate the network nodes' mobility based on the group meetings defined in the previous sections. GRM mobility is inspired by the SWIM mobility model~\citep{swim}. However, instead of defining the nodes' trajectories based on individual decisions, the group defines its meeting places to provide common benefit to its members. 

As in SWIM, GRM defines a home for each node with uniform probability. Then the simulation space is divided in equally sized square cells, and each group $G_i$ assigns to each cell $C_j$ a weight $W(C_j,G_i)$, which is proportional to the average distance of that cell to the homes of each of the members of $G_i$:

\begin{equation}
\begin{split}
W(C_j,G_i)= \frac{1}{||G_i||} \sum_{}^{U_k \in G_a}{\emph{dist}(\emph{Home\/}(U_k),C_j)} \end{split}
\end{equation}

Similarly to the SWIM model, in GRM the \emph{dist\/} function has power-law decay with the euclidean distance, which enables the generation of truncated power-law flights in the users displacements~\citep{individual_mob}. Finally, each cell $C_j$ receives a probability of hosting the group $G_i$ meeting as:
\begin{equation}
\begin{split}
P_{\rm{place}}(C_j,G_i) = \frac{W(C_j,G_a)}{\sum_{i=0}^{N_{\rm{cells}}}{W(C_i,G_a)}}
\end{split}
\end{equation}
where $N_{\rm{cells}}$ denote the total number of cells in the model space.

In GRM, nodes transition between their homes and their group meetings. If the next group meeting is to happen before the necessary time for a node to arrive at home, nodes transition directly between the two meeting places.

\subsection{Automatic Parameters Extraction}\label{auto}

As we have discussed throughout Sec.\ref{grm}, properties such as inter-contact time, contact duration, and groups' sizes in real-world mobility traces are known to follow truncated power laws. However, the parameters (in our notation: $\alpha_{\rm{gmt}}$, $\beta_{\rm{gmt}}$, $\alpha_{\rm{dur}}$, $\beta_{\rm{dur}}$, $\alpha_{\rm{size}}$, $\beta_{\rm{size}}$) of such probability distribution functions vary in different scenarios. For example, contact duration, ICT, and group sizes in a city and in a university campus both follow truncated power laws, but the parameters of the power laws have different values.

GRM allows the user to set these parameters manually, based on previous characterizations of mobility scenarios. However, to ease such parameter extraction, especially in the case when statistical parameters are not known for a given scenario, GRM also supports automatic parameter extraction. For the latter use case, GRM must receive a mobility trace as input.

The automatic parameter extraction is implemented using Maximum Likelihood Estimation (MLE) to find the statistical parameters that have the best fit to the mobility trace given as input. This is a useful feature, because it gives flexibility to GRM usage. It allows the generation of traces containing thousands of nodes based on smaller real-world traces while preserving the same mobility patterns. Hence, network protocols can be tested not only with respect to their performance in real-world traces, but also with respect to their scalability, i.e., their performance under the same scenario but with a higher number of users. Since scalability is an important issue that is often disregarded in opportunistic networking evaluation, due to the lack of large-scale traces, we hope that this feature will allow more comprehensive performance assessments.

\section{Evaluation}\label{eval}

\subsection{Statistical Properties}

A good mobility model must represent well fundamental statistical properties of real-world mobility that are within its scope. In this section, we show that mobility traces generated by GRM maintain the typical characteristics of real mobility that are fundamental for mobile opportunistic networking protocols.

The first properties we evaluate in GRM are pairwise inter-contact time and contact duration. Inter-contact time measures the time between the contacts of pairs of nodes. This is important because, in mobile networks, these contacts are the opportunities to forward messages to other nodes. Conversely, contact duration is important because it determines the amount of data that can be transferred during a given contact. Several studies \citep{swim,ChaintreauTMC2007,leguay2006opportunistic} have used a wide number of real-world traces to show that the inter-contact time and contact duration in human mobility distributions follow truncated power laws.

\begin{figure}[!t]
\centering
    \subfigure[Inter-contact time CCDF]{\includegraphics[width=0.49\columnwidth]{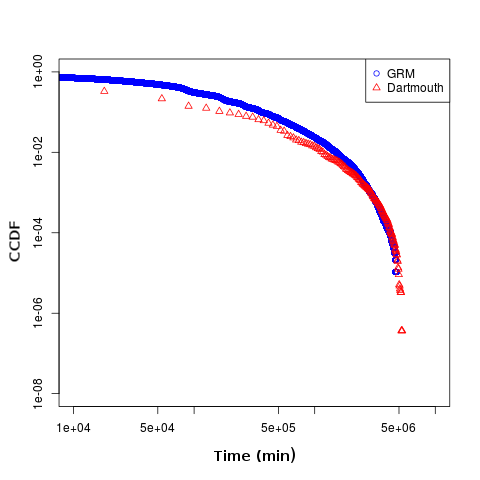}\label{ict_grm}}
    \subfigure[Contact Duration CCDF]{\includegraphics[width=0.49\columnwidth]{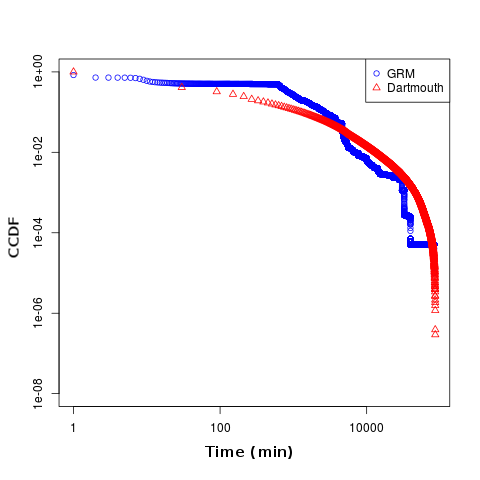}\label{cd_grm}}
    \subfigure[Group Meetings Regularity]{\includegraphics[width=0.49\columnwidth]{best_periodicity.png}\label{reg_grm}}
    \subfigure[Community structure (each color represents a different community)]{\includegraphics[width=0.49\columnwidth]{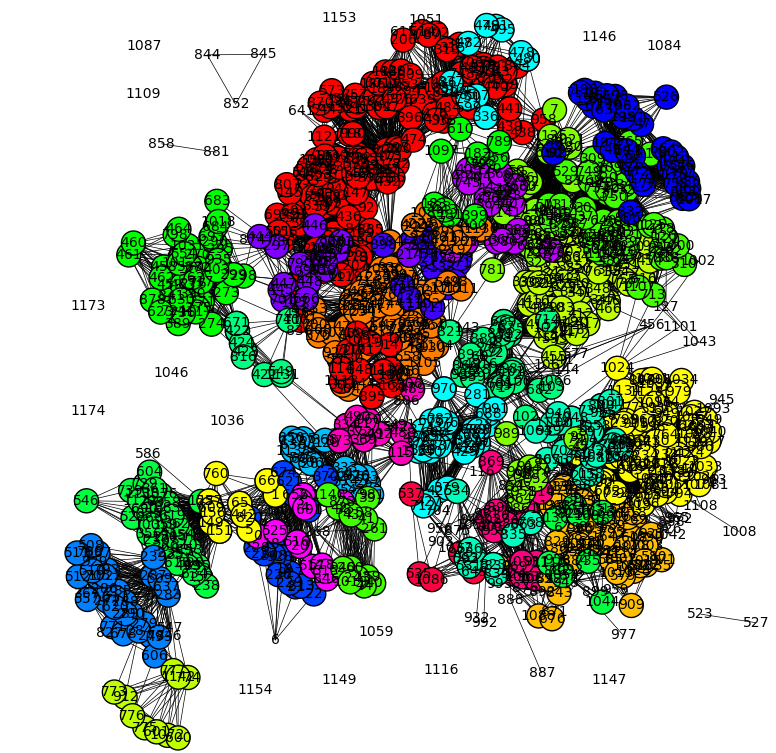}\label{comm}}
\caption{Important properties for opportunistic forwarding extracted from GRM.}
\end{figure}

Fig.~\ref{ict_grm} compares the distribution of inter-contact times for GRM and Dartmouth traces. We see that the inter-contact time distribution of GRM conforms with the one presented in the Dartmouth trace. Both of them follow power laws with exponential cut-offs, also conforming with the results for real-world mobility reported in previous studies. In Fig.~\ref{cd_grm}, we see that the contact duration distribution also follows a power law, conforming with the distributions shown in real human mobility. 

Fig.~\ref{reg_grm} shows that GRM indeed simulates well the regularity of group meetings. We see that the distribution of group re-meeting times is very similar to the ones of real mobility traces (recall Figs~\ref{peri_mit} and~\ref{peri_dartmouth}). It presents peaks at periods of 24 hours and 7 days, remarking the presence of daily and weekly periodicity. This result confirms that GRM fulfill its purpose of properly modeling the role of group meetings regularity in human mobility.

Finally, Fig.~\ref{comm} presents a very important result. It illustrates communities detected in the GRM trace using the Clique Percolation Method~\citep{palla2005}. Such result confirms that, by generating regular group meetings, composed of members who share social bonds (defined in the social network input), the social community structure emerges naturally in the mobile network. Therefore, the traces generated by GRM also account for the influence of social context in human mobility.

\subsection{Opportunistic Forwarding Using GRM Synthetic Traces}\label{forwarding}

One of the most common use cases for mobility models is in the evaluation of mobile networking protocols. Therefore, a mobility model should ideally mimic well the behavior of such protocols in real-world mobility traces. In this section, we evaluate how state-of-art protocols for opportunistic networking perform on GRM. Since the socially-aware strategies are remarkably the most successful for this type of network, this is the focus of our analysis. Namely, we evaluate the performance of Flooding, Bubble Rap~\cite{bubble}, and Groups-Net~\cite{wcm} strategies.

Flooding, which is also referred to as Epidemic Propagation, works by always forwarding a message whenever a node that has such message encounters a node that does not have it yet. In large-scale networks, flooding is not a practical solution, because the number of message re-transmissions grows exponentially, generating a huge network overhead. However, it establishes the upper bound for the messages' delivery ratio and for the network overhead, i.e., no other protocol can deliver more messages or re-transmit more than Flooding.

The Bubble Rap algorithm identifies social communities by looking at densely interconnected nodes in the aggregated contact graph of the trace. Each node in the network must belong to at least one community. Nodes that do not belong to any community are assigned to a pseudo-community of one node. This is necessary for the forwarding algorithm operation. Also, each node gets a measure of its global popularity in the network (\emph{GlobalRank}) and a local measurement of popularity, which is valid within that node's community (\emph{LocalRank}). Using these parameters, the forwarding strategy works as follows. At each encounter, a given node transmits its content if the encountered node has a higher \emph{GlobalRank}, or if the encountered node belongs to a community of which the final destination is a member. Once the message is inside the final destination's community, the forwarding process occurs if the \emph{LocalRank\/} of the encountered node is higher than the \emph{LocalRank} of the node that has the message. This procedure goes on until the message reaches the destination.

The Groups-Net algorithm works by forwarding messages from the origin node to the destination node through the most probable group-to-group path. To define the most probable group-to-group path, the algorithm considers the probability of groups of nodes re-meeting in the near future and the probability of a message being carried between two different groups by a person who is a member of both groups. The probability for a group to meet again in the future is defined based on the assumption that groups that have met more times in the recent past have a greater chance of meeting again in the near future. Since Groups-Net is a group regularity based forwarding policy, it makes perfect sense to evaluate it in GRM.

In the evaluation we consider the following traditional metrics:
\begin{glist}{$\bullet$}{0}{0mm}
\topsep 0mm
\parskip 0mm
\partopsep 0mm
\parindent 0mm
\itemsep 0mm
\parsep 0mm
\item \textbf{Delivery ratio}: Evaluates the percentage of successfully delivered messages for different values of message Time To Live (TTL).
\item \textbf{Number of transmissions}: Measures the network overhead, i.e., the number of device-to-device transmissions that each algorithm performs for different TTLs.
\end{glist}

Opportunistic forwarding algorithms usually have the goal of providing cost-effective message delivery, i.e., the highest possible delivery ratio with the lowest possible network overhead.

\begin{table}[]
\centering
\caption{Simulation parameters}
\label{params}
\begin{tabular}{|c|cc|}
\hline
\multirow{2}{*}{Parameter}           & \multicolumn{2}{c|}{Scenarios}                     \\ \cline{2-3} 
                                     & GRM-100                 & GRM-1000                 \\ \hline \hline
\# of Nodes                              & 100                     & 1000                     \\ 
\# of Groups                             & 500                     & 5000                     \\ 
Sim.\ duration                        & \multicolumn{2}{c|}{60 days}                       \\ 
Groups' durations                      & \multicolumn{2}{c|}{30 days}                       \\ 
Grid                                 & \multicolumn{2}{c|}{30 x 30}                       \\  
Cell size                            & \multicolumn{2}{c|}{50}                            \\  

$\alpha_{\rm{gmt}}$                       & \multicolumn{2}{c|}{2}                             \\  
$\beta_{\rm{gmt}}$                        & \multicolumn{2}{c|}{30 days}                       \\ 
$\alpha_{\rm{dur}}$                       & \multicolumn{2}{c|}{2}                             \\  
$\beta_{\rm{dur}}$                        & \multicolumn{2}{c|}{30 days}                       \\  
$\alpha_{\rm{size}}$                      & 2.24                     & 2.42                    \\ 
$\beta_{\rm{size}}$                       & 30                       & 50                      \\  
K                                    & \multicolumn{2}{c|}{70\%-24h; 15\%-7days; 15\%-6h} \\ 
\multicolumn{1}{|l|}{Social Network} & \multicolumn{2}{c|}{Gaussian Random Partition~\cite{brandes2003experiments}}     \\ \hline
\end{tabular}
\end{table}

To evaluate the algorithms in GRM, we have generated two experimental simulation scenarios containing 100 and 1000 nodes. The simulation parameters for each scenario are specified in Table~\ref{params}.

\begin{figure}[!t]
\centering
  \subfigure[Delivery (100 nodes)]{\includegraphics[width=0.49\columnwidth]{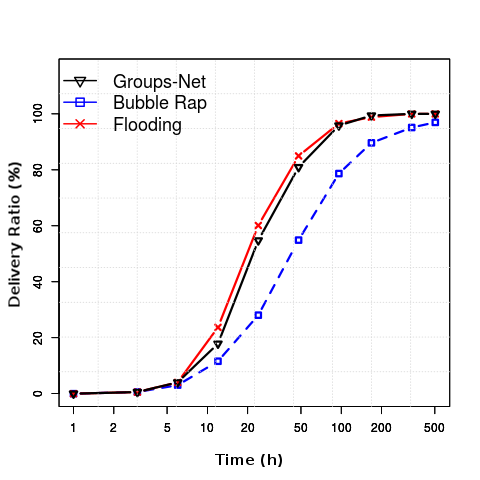}\label{grm_deliv}}
  \subfigure[Overhead (100 nodes)]{\includegraphics[width=0.49\columnwidth]{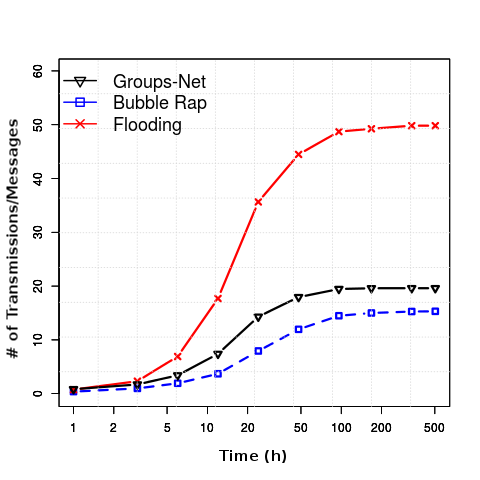}\label{grm_overhead}}
  \subfigure[Delivery (1000 nodes)]{\includegraphics[width=0.49\columnwidth]{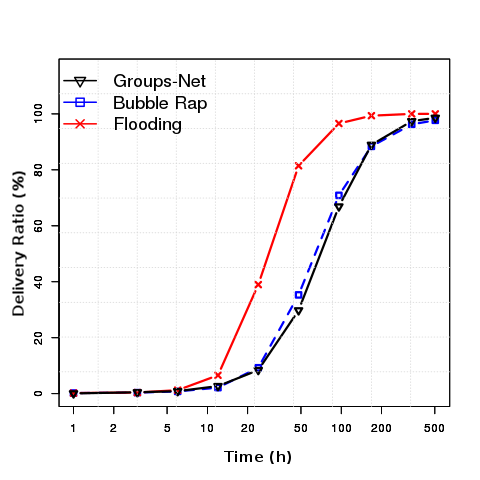}\label{grm_deliv1000}}
  \subfigure[Overhead (1000 nodes)]{\includegraphics[width=0.49\columnwidth]{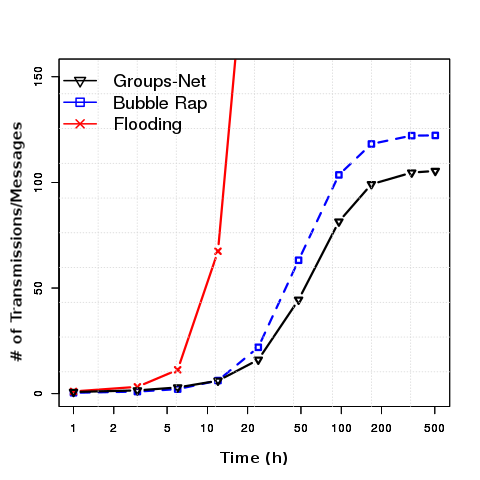}\label{grm_overhead1000}}  
\caption{Flooding and Bubble Rap performances in GRM synthetic trace with 100 nodes}\label{performance}
\end{figure}

Fig.~\ref{performance} presents the performance of the three considered protocols in synthetic traces generated by the GRM model. The result shows that the performances of such algorithms in GRM conforms with their performances in real mobility traces, as reported in the original studies~\cite{wcm,bubble}. They present high delivery ratios, which are comparable to the flooding delivery. On the other hand, by exploring the social context, in the form of communities in Bubble Rap, and in the form of group meetings awareness in Groups-Net, such protocols provide low network overhead, since they only forward messages to the appropriate nodes.

With the increase in the number of nodes in the network, from 100 to 1000, we see that the flooding overhead grows extremely fast, as expected due to its promiscuous forwarding policy. In the trace with 100 nodes, Bubble Rap presents a lower overhead than Groups-Net. When the number of network nodes increases, the overhead of Groups-Net becomes smaller than the Bubble Rap's overhead. As discussed in~\cite{wcm}, the Bubble Rap's overhead presents a linear increase with the number of nodes in the network. This behavior is explained by the greedy nature of Bubble Rap's algorithm. On the other hand, Groups-Net's overhead remains stable, improving its performance in large-scale scenarios.

\section{Conclusion}\label{conclusion}

In this work, we have designed and evaluated GRM, a novel mobility model to represent group meetings regularity and its impact on human mobility. We show that GRM preserves the properties of human mobility that are fundamental for opportunistic networking, namely, ICT and Contact Duration distributions, social community structures, and group meetings regularity. Moreover, we show that GRM mimics well the performance of opportunistic forwarding protocols in real mobility traces, including group meetings-based forwarding. Mobility traces generated by GRM as well as its source code are publicly available.

The existence of a representative social mobility model, such as GRM, enables several opportunities for future research in social-aware forwarding for D2D opportunistic networks. For instance, it would be interesting to evaluate how the existing forwarding schemes perform in large-scale networks with thousands of network nodes. Finally, it is worth mentioning the possibility of extending GRM to incorporate other mobility features, such as map-based trajectories or different visitation frequencies for different regions (Points of Interests), in the trace space. 

Since GRM traces and source code are publicly available, we expect them to be of good use to the mobile networking and mobility modeling research communities. We are hopeful that mobile networking protocol designers will consider GRM in their protocols' evaluation.




\end{document}